# Intensity oscillations observed with Hinode near the south pole of the Sun: leakage of low frequency magneto-acoustic waves into the solar corona

A.K. Srivastava [1, *, #], D. Kuridze [2], T.V. Zaqarashvili [2] & B.N. Dwivedi[3]

1. *Armagh Observatory, College Hill, Armagh BT61 9DG, N.Ireland.*

2. *Abastumani Astrophysical Observatory at I. Chavchavadze State University, Al Kazbegi ave. 2a, 0160 Tbilisi, Georgia.*

3. *Department of Applied Physics, Institute of Technology, Banaras Hindu University, Varanasi-221005, India.*

  * Present Address : ARIES, Manora Peak, Nainital- 263 129, India.
  # Email : aks@aries.ernet.in

**Abstract.**
*Aims.* To study intensity oscillations in the solar chromosphere/corona above a quiet-Sun magnetic network.
*Methods.* We analyse the time series of He II 256.32 Å , Fe XI 188.23 Å and Fe XII  195.12 Å spectral lines observed near the south pole by EUV Imaging Spectrometer (EIS) on board Hinode. We then use a standard wavelet tool to produce power spectra of intensity oscillations above the magnetic network.
*Results.* We find  ~7  min intensity oscillations in all spectral lines and ~13 min intensity oscillation only in He II with the probability of ~96-98 %, which show the most likely signature of magneto-acoustic wave propagation above the network.
*Conclusions.* We suggest that field-free cavity areas under bipolar  magnetic canopies in the vicinity of magnetic network may likely serve as resonators for the magneto-acoustic waves. The cavities with photospheric sound speed and granular dimensions may produce waves with observed periods. These waves may propagate upwards in the transition region/corona and cause observed intensity oscillations.

## 1. Introduction

Complex magnetic field is crucial to influenceing the dynamics of the solar atmosphere. It may channel mechanical energy from the photosphere upwards and may also lead to various wave processes such as reflection, channeling or conversion depending on the field  structure and strength. Vertical magnetic tubes may support the  propagation of tube (e.g., kink, sausage, torsional) waves, while inclined  magnetic fields may cause wave reflection/conversion (Rosenthal et  al. 2002) and/or channeling (De Pontieu et al. 2004).

Observations support the influence of magnetic field on the dynamics of waves. For instance, It has been shown that the quiet-Sun magnetic network elements are surrounded by "magnetic shadows", which lack the oscillatory power at higher frequency range (McIntosh & Judge 2001; Krijger et al. 2001; Vecchio et al. 2007). The magnetic shadows probably correspond to so called "acoustic halos", which are the places of increased high-frequency (>3.3 mHz) power in the photosphere (Muglach et al. 2005). Thus the high-frequency acoustic waves are somehow reflected (or partially conversed) in the chromosphere. On the other hand, lower-frequency (< 3.3 mHz) oscillations are intensively observed over the magnetic network (McAteer et al. 2002, Vecchio et al. 2007), which indicates their penetration in the upper regions. The magnetic field topology is most probable reason of the "wave filtering". The magnetic field is probably vertical in the network cores, but it has a small-scale close structure in the vicinity (Schrijver & Title 2003). Recent observations also show the small-scale bipolar magnetic fields above granular cells (de Wijn et al. 2005; Centeno et al. 2007). Recently, Kuridze et al. (2007) have studied the influence of this small-scale magnetic canopy on wave dynamics. They found that the field-free cavity regions under the canopy may trap high-frequency acoustic oscillations, while the lower-frequency oscillations may be channeled upwards in the form of magneto-acoustic waves.

In this paper, we study the spectrum of intensity oscillations above the quiet-Sun magnetic network using the new data from EIS on board Hinode. The observational results are then interpreted by magnetohydrodynamic (MHD) modeling of wave dynamics in a cylindrical bipolar magnetic canopy.

## 2. Observations and data reduction

We use the time series data of on-disk quiet Sun near the south pole as observed by 40"-slot of EIS (Culhane et al. 2006). The 40" and 266" slots are for image analyses using light curves, while 1" and 2" slits are for spectral and Doppler analyses using spectral line profiles. EIS observes high resolution spectra in two wavelength bands 170-211 Å and 246-292 Å by short-wavelength (SW) and long-wavelength (LW) CCDs respectively. The spectral resolution of EIS is 0.0223 Å per pixel. There are *approximately* 16 pixels and 2 pixels offsets respectively in the Y and X-coordinates of these two CCDs (Young et al. 2007). The observational sequence was taken on 11th March, 2007, and the name of study is HPW005_QS_Slot_60m. The data contain He II 256.32 Å, Si VII 257.35 Å, Fe VIII 185.45 Å, Fe X 184.54 Å, Si X 258.37, 261.04 Å, Fe XI 188.23 Å, Fe XII 188.86, 195.12 Å, Fe XIII 202.04 Å, Fe XIV 264.78, 274.2 Å, Fe XV 284.12 Å, Fe XVI 262.92 Å, Ca XVII 192.82 Å spectral lines. The observation started on 18:12:33 U.T. and ended at 19:04:06 U.T. The position of the X-centre and Y-centre of slot are respectively 110" and -973", while the X-FOV and Y-FOV are respectively 40" and 512" (Fig. 1). The binning of data is 1" x 1". The exposure time for each spectral line is 60 s. The integration time for each step of time series is uniform, being 60 s. We choose the brightest core of quiet-Sun magnetic network in three different spectral lines: He II 256.32 Å, Fe XI 188.86 Å and Fe XII 195.12 Å. He II is observed by LW CCD [$(X_{LW}, Y_{LW})$ = ($36^{th}$ pixel, $46^{th}$ pixel)], while the iron lines are observed by SW CCD. Therefore, we can *approximately* compensate the offset of iron lines by following relation: $X_{SW} = (X_{LW} -2)$, $Y_{SW} = (Y_{LW} + 16)$.

We start with raw (zeroth level) data and use the standard EIS subroutines for calibration. The subroutines can be found in the SSWIDL software tree (e.g., www.Darts.isas.jaxa.jp/pub/solar/ssw/hinode/eis/). These standard subroutines correct for dark current subtraction, cosmic ray removal, flat field correction, hot pixels, and bad/missing pixels. The data is saved in the level-1 data file, while associated errors are saved in the error file. S/N ratios of chosen lines (He II 256.32 Å , Fe XI 188.23 Å, Fe XII 195.12 Å) lie between 20-50 which is good for real analyses. The summation of X or Y pixels has not been carried out to prevent the information losses about small bright features in the magnetic network.

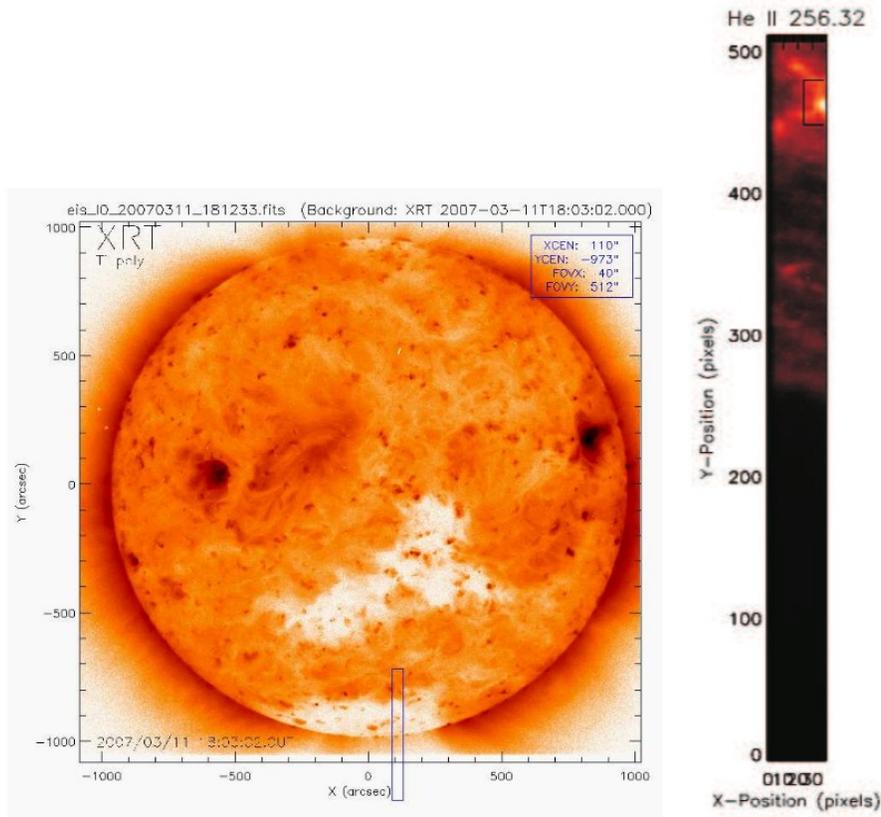

**Fig 1.** *The pointing position of 40" slot over the full disk image of XRT/Hinode (left), and the brightened magnetic network is shown in He II 256.32 Å line (right). The position on the full disc image is expressed in arc sec and the position on the slot in pixels.*

## 3. Results

We use the Morlet wavelet tool to produce the power spectrum of oscillations in different spectral lines. We note that the Morlet wavelet suffers from the edge effect of time series data. However, this effect is significant in regions defined as cone of influence (COI). The details of wavelet procedure, its noise filtering, COI effects etc. are given in Torrence & Compo (1998).

Randomisation technique evaluates the peak power in the global wavelet spectrum, which is just the average peak power over time and similar to a smoothed Fourier power spectrum. This technique compares it to the peak powers evaluated from the n! equally likely permutations of the time series data, assuming that n values of measured intensities are independent of n measured times if there is no periodic signal. The proportion of permutations which gives the value greater or equal to the original peak power of the time series, will provide the probability of no periodic component (p). The percentage probability of periodic components presented in the data will be (1-p) x100, and 95 % is the lowest acceptable probability for real oscillations. We set 200 permutations for the reliable estimation of p, and hence the probability of real oscillations. The details of randomisation technique to obtain the statistically significant real oscillation periods are given by Nemec & Nemec (1985) and O'Shea et al. (2001). We do not remove any upper/longer period intervals and associated powers of our time series data during wavelet analysis. However, we do not consider first 3-4 min in analyses as this is associated with the abrupt intensity enhancement. We choose the 'running average' option of O'Shea's wavelet tool, and smooth the original signal by window of scalar-width 10. Average smoothing process is used to reduce the noise in the original signal in order to get a real periodicity, and it is based on the low pass filtering methods. The maximum allowed period from COI, where edge effect is more effective, is 1039 s. Hence, the power reduces substantially beyond this threshold. In our wavelet analysis, we only consider the power peaks and corresponding real periods below this threshold. The results of the wavelet analysis for the spectral lines He II 256.32 Å, Fe XI 188.23 Å and Fe XII 195.12 Å are given in Figs. 2-4. We get the periodicity of ~ 6.5-7.5 min with the probability of 96-98 % in all spectral lines and additionally ~ 13 min periodicity in He II line. It should be noted that ~ 7.5 min and ~ 13 min, both the periodicities are statistically significant in chromospheric He II line. These oscillations are likely signature of the propagation of low-frequency magneto-acoustic waves near the magnetic network. We have also performed the wavelet analysis in non-network and non-brightened region. We have chosen the coordinate ($X_{sw}$,$Y_{sw}$) ~ (26, 326) for iron lines and corresponding offset compensated coordinates for the He II line. The wavelet gives the periodicity of ~ 13-16 min with the probability between 77 % and 83 %, which is not a real oscillation according to O'Shea et al. (2001). Hence, ~ 7 and ~ 13 min oscillations are only apparent in the brightened magnetic core region.

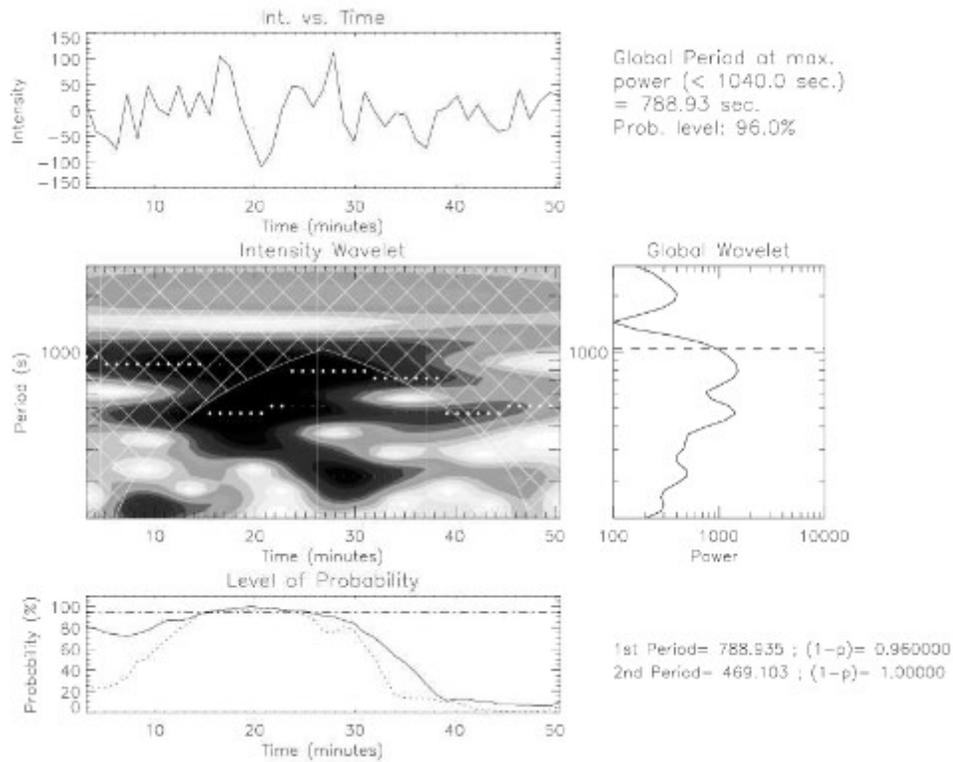

**Fig 2.** *The wavelet result for He II 256.32 Å (top) line: The top panel shows the variation of intensity, the wavelet power spectrum is given in the middle panel, and the probability is given in the bottom panel.*

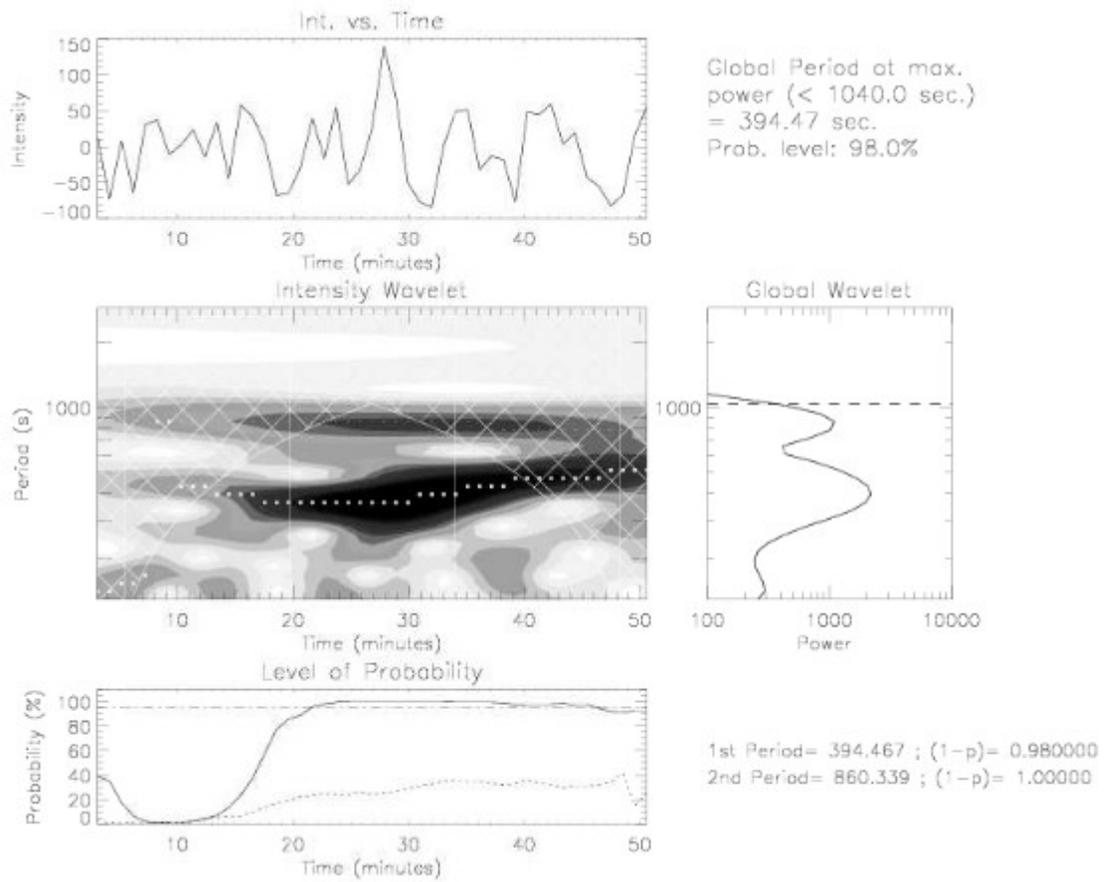

**Fig 3.** *The wavelet result for Fe XI 188.23 Å (top) line : The top panel shows the variation of intensity, the wavelet power spectrum is given in the middle panel, and the probability is given in the bottom panel.*

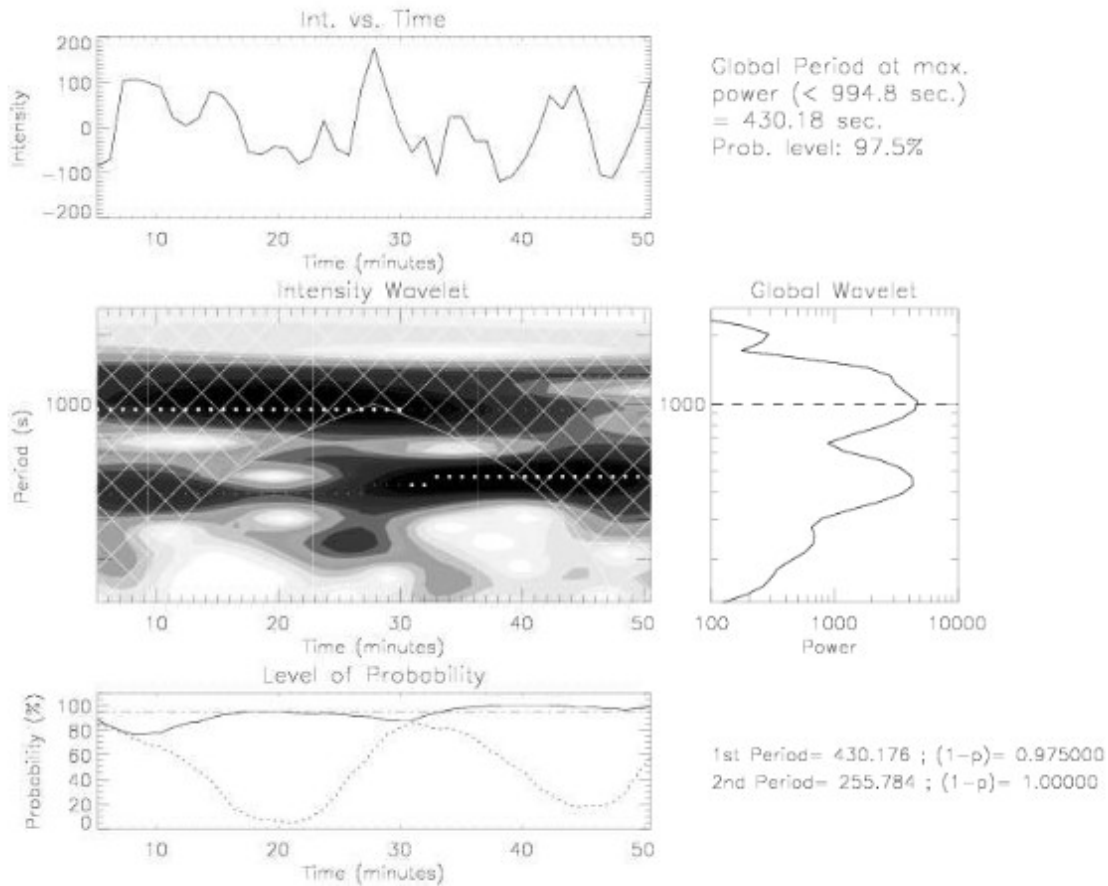

**Fig 4.** *The wavelet result for Fe XII 195.12 Å (top) line: The top panel shows the variation of intensity, the wavelet power spectrum is given in the middle panel, and the probability is given in the bottom panel.*

## 4. A theoretical model

As already noted, Observed intensity oscillations in different spectral lines over the magnetic network likely carry the signature of the propagation of magneto-acoustic waves from the photosphere. However, observed frequencies are below the photospheric cut-off value. Therefore, the waves should be evanescent in the lower atmosphere for purely vertical propagation. Consequently, their observation in the higher atmosphere provides compelling ground for debate. These can be resolved if the waves are either generated i*n situ* or they propagate at an arbitrary angle which would reduce the cut-off frequency (De Pontieu et al. 2004). Here, we present a theoretical model which supports the wave propagation from the photosphere upwards.

The magnetic field is probably vertical in the core of network, but becomes inclined in the vicinity as small-scale bipolar loops (McIntosh & Judge 2001; Schrijver & Title 2003). Field-free cavities of granular dimensions under this small-scale cylindrical canopy may become effective resonators for MHD waves. Higher order harmonics may be trapped in the cavity but the first harmonic may propagate upwards. The photospheric cut-off frequency for all harmonics will be lower in comparison to the actual cut-off value, which is due to a finite wave number component in the horizontal direction.

Cylindrical bipolar magnetic field, similar to McIntosh & Judge (2001) and Schrijver & Title (2003), is used to model wave behaviour in the vicinity of magnetic network. We suggest that a field-free cavity area ($r<r_0$ region in Fig. 5) overlayed by the magnetic canopy can be formed over granular cells when plasma flows transport the magnetic flux at boundaries (Centeno et al. 2007). The magnetic field, $B_0$ in the overlying canopy ($r>r_0$ region in Fig. 5) is suggested to be current-free with only a ø-component in a cylindrical coordinate system ($r,ø,z$), i.e. **B₀**=$(0, B_ø(r), 0)$, where $B_ø(r)=B_{ø0}r_0/r$ (Díaz et al. 2006).

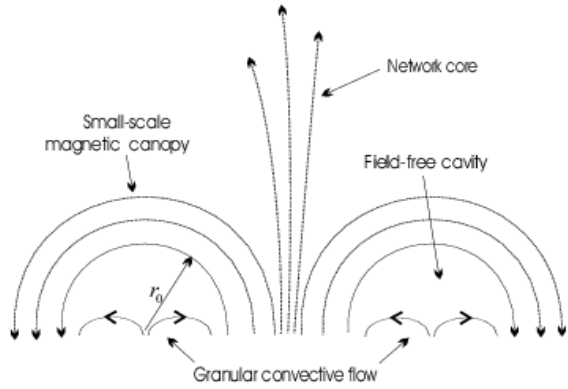

**Fig 5.** *Schematic view of a small-scale bipolar magnetic canopy overlying a field-free cavity region near a network core.*

The oscillation spectrum can be obtained by solving governing plasma equations in both, the canopy and the cavity regions separately. Then the merging of the solutions at canopy/cavity interface gives the dispersion relation for oscillations. Fourier analysis of linearized adiabatic hydrodynamic equations with respect to both the time (*t*) and coordinate (*ø*) leads to the Bessel equation in the field-free cavity region:

$$\frac{\partial^2 \rho_1}{\partial r^2} + \frac{1}{r}\frac{\partial \rho_1}{\partial r} + \left[\frac{\omega^2}{c_0^2} - \frac{m^2}{r^2}\right]\rho_1 = 0, \qquad (1)$$

where $m$ corresponds to the azimuthal wave number, $\rho_1$ is the density perturbations, and $c_0=(\gamma p_0/\rho_{01})^{1/2}$ is the adiabatic sound speed. Here $p_0$ and $\rho_{01}$ are unperturbed pressure and density in the cavity ($r<r_0$) and $\gamma$ is the ratio of specific heats. Solution of this equation is the Bessel function $J_m(k_1 r)$, where $k_1=\omega/c_0$.

In the magnetic canopy region ($r>r_0$), we use the cold plasma approximation. In this case, linearised MHD equations lead to:

$$\frac{\partial^2 \xi_r}{\partial r^2} - \frac{1}{r}\frac{\partial \xi_r}{\partial r} + \left[\frac{\omega^2}{v_A^2(r)} + \frac{1}{r^2} - \frac{m^2}{r^2}\right]\xi_r = 0, \qquad (2)$$

where $\xi_r$ is the transverse Lagrangian position vector component and $v_A = B_\phi / (4\pi \rho_{02})^{1/2}$ with $\rho_{02}$ being unperturbed density inside the canopy region. The solution of this equation is proportional to the Hankel function of half integer order $rH_{m/2}(k_2 r)$, where $k_2=\omega/2v_A$ (Díaz et al. 2006).

The continuity of the transverse Lagrangian position vector component and the total Lagrangian pressure perturbation at the canopy/cavity interface $r=r_0$ leads to the general dispersion relation (Kuridze et al. 2007):

$$\frac{\omega^2}{v_A^2}\frac{\rho_{01}}{\rho_{02}}\frac{J_m(k_1 r_0)}{k_1 J'_m(k_1 r_0)} = -\frac{k_2 r_0 H'_{m/2}(k_2 r_0)}{r_0 H_{m/2}(k_2 r_0)}. \qquad (3)$$

Eq. (3) describes the oscillation spectrum in the cavity region under the magnetic canopy (Fig. 5). The wave frequency is generally complex, which means the leakage of oscillations upwards. Kuridze et al. (2007) show that higher order $m>1$ harmonics are trapped in the cavity, while the first $m=1$ harmonic is leaky. This means that the first harmonic may propagate upwards as the fast magneto-acoustic wave and cause the intensity oscillations in coronal spectral lines. Therefore, we emphasize on the behaviour of the $m=1$ harmonic in this context.

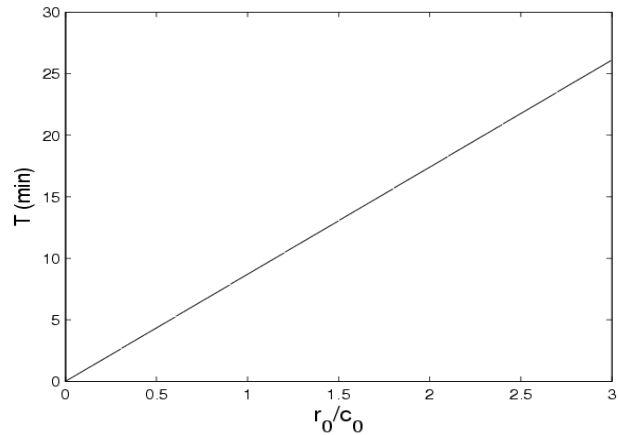

**Fig 6.** *The period of m=1 harmonic vs the ratio of cavity size and sound speed $r_0/c_0$. The $r_0/c_0$ is normalised by 1 min The observed wave period ~ 7 min and ~ 13 min correspond to the radius of ~ 450 km and ~ 800 km for the sound speed of 9 km/s.*

Fig. 6 shows the dependence of the period of $m$=1 harmonic on the ratio of cavity radius $r_0$ and sound speed $c_0$. Hence, the typical photospheric sound speed and the cavity radius determine the oscillation period. We infer from Fig. 6 that the photospheric sound speed $c_0$= 9 km/s gives the observed period of ~ 6.5-7.5 min for the cavity diameter of ~ 800-900 km and ~13 min for the diameter of ~1600 km. This probably indicates to the enhanced power at granular radius of the same scales. Indeed, a photometric and spectroscopic study of abnormal granulation near the strong magnetic fields shows the enhanced power at granular diameter of 1.3" (~900 km) and 2.5" (~1800 km) (Sobotka et al. 1994). Thus, the enhanced power at granular size near the magnetic network core can be responsible for the observed oscillation periods.

## 5. Discussion and conclusions

The analysis of new observational data obtained by EIS on board Hinode show ~7 min intensity oscillations in He II 256.32 Å, Fe XI 188.23 Å and Fe XII 195.12 Å spectral lines with the probability of 96-98 % above the magnetic network near the south pole. In addition, we also found ~ 13 min intensity oscillations in He II 256.32 Å. The oscillations have been observed in three different spectral lines formed at different plasma temperatures (*log $T_{He\ II}$ = 4.90, log $T_{Fe\ XI}$= 6.10, log $T_{Fe\ XII}$= 6.20*). Therefore, the oscillations are probably presented at different heights, which imply the wave propagation from the photosphere into the transition region and corona.

We suggest for the first time that the field-free cavity regions under the bipolar small-scale magnetic canopy in the vicinity of network core may serve as resonators for the waves. This magnetic field configuration has been suggested to be the case near the network core (McIntosh & Judge 2001; Schrijver & Title 2003) and also above granular cells (de Wijn et al. 2005; Centeno et al. 2007). The field-free cavities under the bipolar magnetic canopy probably have granular dimensions as the granular flow transport the magnetic flux at cell boundaries. We have solved the MHD equations in the cavity and canopy regions separately and merged the solutions at cavity-canopy interface. Consequently, we have obtained the dispersion relation for oscillations, which shows that the first harmonic is leaky thus the oscillations may propagate upwards. The period of the first harmonic depends on the cavity radius and the sound speed. In order to get the periods of ~ 7 and 13 min, the cavity diameter should be correspondingly ~ 900 and 1600 km for the photospheric sound speed of ~ 9 km/s. Surprisingly, this result shows up the power spectra of abnormal granulation, which depicting enhanced power at granular diameters of 900 km and 1800 km (Sobotka et al. 1994). Therefore, observed oscillations can be explained by wave excitation in cavities under the small-scale canopy. It should be mentioned, that the frequency of the first harmonic is above the photospheric cut-off frequency because of the horizontal component of wave vector. Therefore, the oscillation may easily penetrate into the higher regions. However, inclusion of gravitational stratification is necessary for a more clear link between observations and theory.

The oscillations with ~ 7-13 min have been observed only in the brightest core of magnetic network. We could not find statistically significant oscillations in nearby non-magnetic regions. Therefore, it seems that the oscillations are typical for magnetic networks. However, the similar analysis should be carried out above other network elements as well. This is a subject of future detailed analysis of Hinode and ground-based observational data.

## Acknowledgment

AKS thanks Prof. J.G. Doyle for valuable discussions and encouragements, and Dr. E. O'Shea for 'RANDOMLET'. TVZ and DK thank Prof. S. Poedts and Dr. B. Shergelashvili for valuable discussions. This work was supported by the grant of Georgian National Science Foundation GNSF/ST06/4-098. Hinode is a Japanese mission developed and launched by ISAS/JAXA, with NAOJ as domestic partner and NASA and STFC (UK) as international partners. It is operated by these agencies in co-operation with ESA and NSC (Norway). We also acknowledge Dr. Harry Warren for his observing sequence, which is avialable in EIS data archive. We thank to the referee for his valuable suggestions, which considerably improved the manuscript.